# Nash Equilibrium of Joint Day-ahead Electricity Markets and Forward Contracts in Congested Power Systems

Mohsen Banaei
Department of Applied Mathematics
and Computer Science
Technical University of Denmark,
Copenhagen, Denmark
moban@dtu.dk

Majid Oloomi Buygi
*Department of engineering*

*Ferdowsi University of Mashhad,*
Mashhad, Iran
m.oloomi@um.ac.ir

Hani Raouf-Sheybani
*Department of engineering*

*Quchan University of Technology*
Quchan, Iran
hani.raoof@gmail.com

Razgar Ebrahimy
Department of Applied Mathematics
and Computer Science
Technical University of Denmark,
Copenhagen, Denmark
raze@dtu.dk

Henrik Madsen
Department of Applied Mathematics
and Computer Science
Technical University of Denmark,
Copenhagen, Denmark
hmad@dtu.dk

*Abstract*— Uncertainty in the output power of large-scale wind power plants (WPPs) can face the electricity market players with undesirable profit variations. Market players can hedge themselves against these risks by participating in forward contracts markets alongside the day-ahead markets. The participation of market players in these two markets affects their profits and also the prices and power quantities of each market. Moreover, limitations in the transmission grid can affect the optimal behavior of market players. In this paper, a Cournot Nash equilibrium model is proposed to study the behavior of market players in the forward contract market and the day-ahead electricity market in a congested power system with large-scale integration of WPPs. The proposed method is applied to a test system, and the results are discussed.

I. INTRODUCTION

In power systems with large-scale penetration of WPPs, market players are confronted with the financial risks related to the uncertain and uncontrollable output power of these units [1][2]. Variation in the output power of WPPs causes variation in the scheduled power for producers to supply in the power system which means uncertainty in the scheduled power of the producers and suppliers [3]. Moreover, uncertainty in the amount of remained power for producers affects the competition level in the system and consequently the market prices [4]. So, uncertainty in the output power of the WPPs can lead to uncertainty in the scheduled power and electricity prices and consequently uncertainty in the profit of both electricity producers and suppliers. The forward contract is one of the well-known financial derivatives that is used to avoid uncertain profits and obtain stable revenue in the power system. According to the definition, a forward contract is an agreement between a producer and a supplier for delivering a specific amount of energy at a predetermined price at a specific time in the future [5]. Forward contract prices are affected by the expectation of the day-ahead market price in the power delivery period. Day-ahead market prices can also be affected by changing the remained power in the day-ahead market after transferring a portion of total demand to the forward contracts. So, forward contracts and the day-ahead market have mutual impacts on each other and both markets should be considered in order to study the behavior of market players and power system operations.

Forward contracts and day-ahead markets have already appeared in the literature for different viewpoints. The optimal bidding strategy problem in forward contracts and day-ahead market from the viewpoint of a specific producer or supplier is addressed in [6]-[12]. Different methods like dynamic programming and linear programming approaches are used to find optimal solutions. Moreover, the CVaR method is mostly used to address the risk management preferences of the market players. Market regulator viewpoint is addressed [13]-[17], and [5]. The goal of these studies is to determine the Nash equilibrium of the system considering both forward contracts and the day-ahead market. In [13] the Nash equilibrium problem is solved for only one producer and supplier. In [14] a supply function equilibrium model is proposed for the day-ahead market and forward contracts. The strategic behavior of suppliers and congestion in the grid is not included in the model. In [15] the optimal price adjustment problem of forward contracts in a power system with producers and suppliers with flexible and inflexible loads and renewable resources is solved. Reference [16] finds the Nash equilibrium of forward contracts in a power system while the impacts of forward contracts on the price of the day-ahead market are ignored. In [17] and [5] supply function equilibrium models for an electricity market parallel with a forward contract market are proposed. Both Uniform and pay-as-bid pricing mechanisms are considered for the day-ahead electricity market but transmission system constraints are ignored.

In this paper, the behavior of producers and suppliers in forward contracts and day-ahead market in a congested power system with large-scale WPPs is studied. A Cournot Nash equilibrium model is proposed for the joint day-ahead market and forward contracts. The proposed method considers the operational constraints of producers and suppliers, different risk management preferences of market players, transmission system constraints, and uncertainties in the output power of WFs. The rest of the paper is organized as follows: in section II, the problem is defined. In section III, the problem is formulated. Simulation results are analyzed and discussed in section IV, and finally, conclusions are presented in section V.

## I. PROBLEM DESCRIPTION AND ASSUMPTIONS

As mentioned before, forward contracts and day-ahead market have mutual impacts on each other. These impacts can be affected by congestion in the grid. In this paper, a Cournot Nash equilibrium model is proposed to study the effects of congestion in the grid on the equilibrium point of forward contracts and day-ahead market and profit of market players. It is assumed that the producers behave strategically in both forward contracts and the day-ahead market. But, suppliers are strategic market players only in the forward contract negotiations and are price takers in the day-ahead market to cover all their requested demand anyhow. Each producer $i$ is modeled by an aggregated marginal cost function $a_i + b_i Q_{i,s}^{dp}$. $a_i$ and $b_i$ are constant parameters and $Q_{i,s}^{dp}$ is the dispatched power of the producer $i$. Each supplier $j$ is modeled with an aggregated marginal utility function $c_j + d_j Q_{j,s}^{dc}$. $c_j$ and $d_j$ are constant parameters and $Q_{j,s}^{dc}$ is the scheduled power of the consumer $j$. The delivery period is assumed to be one hour on a specific day in the future.

The problem is solved for the contracting period. In this period, market players decide on the quantity and price of their contracts. To find the optimal contract price and quantity, aggregated profit of market players in forward contracts and the day-ahead market is maximized considering their risk management preferences and estimated actions in the day-ahead market for each scenario of WPPs' output power at the delivery period. The structure of the problem is depicted in Fig. 1.

It is assumed the system includes $n_a$ areas. These areas are connected through the power transmission line with limited power transmission capacity. In each area, there is one large-scale supplier and there might be zero, one, or more producers. So, the same index is used for suppliers and areas.

### A. Uncertainty modeling

Wind power generation is considered the main source of uncertainty in the system. The total installed capacity of wind power in each area is modeled as a single WPP with specific characteristics. To create WPPs' output power uncertainty scenarios, first, some discrete scenarios are generated for each area. Then, considering the correlation between the WPP's output power in different areas, one WPP's output power sample e.g., $q_s^{w,k}$ from each area $k$ is extracted [18]. The set $q_s^{w,k}$ for all areas is considered as a scenario for the output power of the WPPs in the grid, i.e., $Q_s^w = \{q_s^{w,1}, q_s^{w,2}, ..., q_s^{w,n_a}\}$ where $Q_s^w$ represents the WPPs' output power in the scenario $s = 1, ..., n_s$.

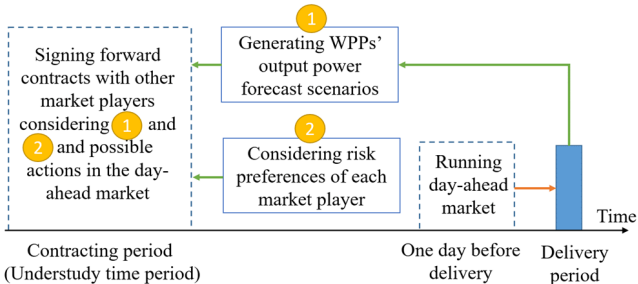

Fig. 1. Problem structure

### B. Proposed risk management method

The proposed risk management method in [5] is used to consider the risk management preferences of the market players in the model. This method is based on the concerns of market players about occurring specific scenarios in the delivery period. For example, increasing (decreasing) the output power of WPPs, reduces (increases) the electricity price in the delivery period, and consequently, reduces the profit of producers (suppliers). So, they are more concerned about the scenarios that lead to increasing (decreasing) the output power of WPPs. According to the concern scenario method, probabilities of uncertainty scenarios for each market player are replaced by "concern values" that reflect the concerns of that market player about happening those scenarios. Assume that we have $Q_s^w \leq Q_{s+1}^w$, for suppliers (producers) we should have $\rho_{j,s}^C \geq \rho_{j,s+1}^C$ ($\rho_{i,s}^P \leq \rho_{i,s+1}^P$) where $\rho_{j,s}^C$ ($\rho_{i,s}^P$) is the concern value for supplier $j$ (producer $i$). It is suggested to use the Exponential probability density function (E-PDF) ($e(x) = \beta e^{-\beta x}$ $x \geq 0$) to generate these concern values for each market player. Fig. 2 shows two sets of extracted discrete concern values from E-PDF for a producer and supplier. As shown in Fig. 2, for suppliers (producers) the first wind power uncertainty scenario is assigned to the first (last) generated concern value, the second scenario is assigned to the second (one before the last) generated concern value, and so on. The parameter $\beta$ can be used to model the amount of concern of market players about the future. A producer $i$ (supplier $j$) with a greater value for $\beta_i^p$ ($\beta_j^c$) represents a more concerned market player [5].

## II. NASH EQUILIBRIUM FORMULATION

To find the Nash equilibrium, first forward contract negotiations are formulated. Then, since the expectations of the day-ahead market are also considered in the contracting period, the day-ahead electricity market is formulated. After that, producers' and suppliers' optimization problems are determined and finally, the Nash equilibrium calculation process is explained.

### A. Forward contracts formulation

It is assumed that each supplier $j$ is allowed to have a contract with each producer $i$ and vice versa. Each producer $i$ (supplier $j$) submits affine bid function $F_{ij}^{fp} = \alpha_{ij}^f + b_i Q_{ij}^{fp}$ ($F_{ij}^{fc} = \varepsilon_{ji}^f - d_j Q_{ji}^{fc}$) to each supplier $j$ (producer $i$). The slope of this function is equal to the slope of the producers' marginal cost function (supplier's marginal utility function) and its intercept i.e., $\alpha_{ij}^f$ ($\varepsilon_{ji}^f$) is the decision-making variable of producer $i$ (supplier $j$) in contract with supplier $j$ (producer $i$). The intersection of these two bids obtains the quantity and price of the contract between producer $i$ and supplier $j$ as below [19]:

$$Q_{ij}^{fp} = Q_{ji}^{fc} = (\varepsilon_{ji}^{f} - \alpha_{ij}^{f})/(b_i + d_j) \quad (1)$$

$$F_{ij}^{fp} = F_{ij}^{fc} = (b_i \varepsilon_{ji}^{f} + d_j \alpha_{ij}^{f})/(b_i + d_j) \quad (2)$$

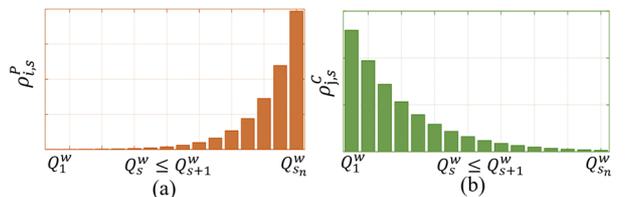

Fig. 2. Concern values for a) producer $i$, b) supplier $j$

$Q_{ji}^{fc}$ and $Q_{ij}^{fp}$ are the agreed quantities of power for a forward contract between producer $i$ and supplier $j$. $F_{ij}^{fc}$ and $F_{ij}^{fp}$ are the agreed prices of the contract between producer $i$ and supplier $j$.

*B. Day-ahead market estimation formulation*

The day-ahead market operation should be formulated for each uncertainty scenario $s$. Cournot model is applied in this paper to model the day-ahead market operation. The proposed formulation method in [20] is upgraded in this paper to model the day-ahead market operation. At [20], first, a bilateral model is proposed. In this model, producers sell their output powers only to the suppliers in their areas. Then, arbitragers purchase energy from low-price areas and sell this energy to high-price areas until the price difference between every two nodes gets equal to the related transmission price. This turns the bilateral model into a pool-co model which can be used for the day-ahead market [21]. In [20] it is also assumed that all produced energy passes through a virtual hub node. The transmission system operator (TSO) charges producers a congestion-based wheeling fee $W_n$ \$/MWh for transmitting power from the hub node to area $n$ and generation scheduling is performed by maximizing the total revenue of the TSO. This proposed pool-co model is upgraded as follows: 1) forward contracts are included in the upgraded model, 2) strategic gaming of suppliers in the forward contracts is modeled, and 3) uncertainties are involved in the model. The TSO optimizations should be solved for each scenario $s$, separately. The optimization problem of the TSO for scenario $s$ of the delivery period is as below:

$$\max \sum_{j \in A} W_{j,s} (\gamma_{j,s} + \sum_{i \in P} Q_{ji}^{fc} - \sum_{i \in P_j} \sum_{i \in C} Q_{ij}^{fp}) \quad (3)$$

$$s.t. \sum_{m \in A} V_j V_m B_{jm} (\delta_{js} - \delta_{ms}) = -(\gamma_{j,s} + \sum_{i \in P} Q_{ji}^{fc} - \sum_{i \in P_j} \sum_{i \in C} Q_{ij}^{fp}) \quad (4)$$
$$\forall j \in A$$

$$V_j V_m B_{jm} (\delta_{js} - \delta_{ms}) \leq \overline{T}_{jm}, \quad \forall (j,m) \in L \quad (5)$$

$$V_j V_m B_{jm} (\delta_{js} - \delta_{ms}) \geq \underline{T}_{jm}, \quad \forall (j,m) \in L \quad (6)$$

$P$ and $C$ are the set of all producers and suppliers. $P_j$ is the set of producers and suppliers in area $j$. $V_j$ is the voltage of area $j$. $B_{jm}$ is the susceptance of the line between area $j$ and $m$. $\delta_{js}$ is the phase angle of area $j$. $\gamma_{j,s}$ is the injected power to area $n$ in the day-ahead electricity market from TSO's viewpoint. DC load flow method is used to model the transmission system operation. $\overline{T}_{jm}$ and $\underline{T}_{jm}$ are the upper and lower bounds of transmission capacity for the line between nodes $j$ and $m$. $L$ is the set of lines. The objective function (3) is the TSO's revenue from transferring power to each area. Constraint (4) is Kirchhoff's current law. Constraints (5) and (6) are the upper and lower transmission capacity of the grid's lines. Since the TSO optimization problem (3)-(6) is solved for the day-ahead market, $Q_{ji}^{fc}$ and $Q_{ij}^{fp}$ are assumed to be constant in (3)-(6).

*C. Producers' profit formulation*

Producers should maximize the sum of their profit in forward contracts and the day-ahead market. So, the optimization problem of producers is as follows:

$$\max E(P_i) = \sum_{s \in S} \rho_{i,s}^{P} [\lambda_{j(i),s} Q_{i,s}^{dp} + \sum_{j \in C} F_{ij}^{fp} Q_{ij}^{fp} \\ - a_i (Q_{i,s}^{dp} + \sum_{j \in C} Q_{ij}^{fp}) - 0.5 b_i (Q_{i,s}^{dp} + \sum_{j \in C} Q_{ij}^{fp})^2] \quad (7)$$

$$s.t. \; Q_{i,s}^{dp} + \sum_{i \in C} Q_{ij}^{fp} \leq \overline{Q}_i, \quad (\overline{\mu}_{i,s}^{p}) \quad \forall s \in S \quad (8)$$

$$\lambda_{j(i),s} = c_j - d_j (\sum_{i \in P_{j(i)}} Q_{i,s}^{dp} + \sum_{i \in P} Q_{ji}^{fc} + x_{i,j(i),s}) \quad \forall s \in S \quad (9)$$

$$\lambda_{j(i),s} = \lambda_{hub,s} + W_{j,s} \quad (\mu_{i,j,s}^{hub}) \quad \forall s \in S, \forall j \in A \quad (10)$$

$$\sum_{j \in A} x_{i,j,s} = 0 \quad (\mu_{i,s}^{x}) \quad \forall s \in S \quad (11)$$

$$Q_{ij}^{fp} \geq 0 \quad (\mu_{ij}^{fp}) \quad \forall j \in C \quad (12)$$

$$Q_{i,s}^{dp} \geq 0 \quad \forall s \in S \quad (13)$$

$\lambda_{hub,s}$ and $\lambda_{j(i),s}$ are hub node price and area $j$ price, respectively. $j(i)$ represents the index of the area $j$ that producer $i$ is located in. $x_{i,j,s}$ is the injected power to area $j$ from the hub node from the producers' viewpoint.

The first and second terms of the objective function (7) are the revenue in the day-ahead market at uncertainty scenarios and the revenue of forward contracts. The last two terms represent the operation cost of the producer. Constraint (8) limits the output power of the producer to the maximum power generation capacity of its units. Constraints (9) represent the Locational Marginal Price (LMP) of area $j$. Constraint (10) presents the relationship between the area prices and the hub node price. Constraint (11) considers the fact that the sum of injected power to all areas from the viewpoint of producer $i$ is equal to zero. Constraints (12) and (13) guarantee the positivity of generated power by producer $i$ in the day-ahead market and forward contracts, respectively. The producer $i$ decision-making variables in forward contracts are $\alpha_{ij}^{f} \; \forall j \in C$ and in each uncertainty scenario of the day-head market is submitted power bid to the market, i.e., $Q_{i,s}^{dp}$.

*D. Supplier's profit formulation*

The profit of suppliers can be formulated by subtracting the utility of the electricity for that supplier from the payment for the electricity. So, we have:

$$\max E(U_i) = \sum_{s \in S} \rho_{j,s}^{C} [-\lambda_{j(i),s} Q_{j,s}^{dc} - \sum_{i \in P} F_{ji}^{fc} Q_{ji}^{fc} \\ + c_j (Q_{j,s}^{dp} + \sum_{i \in P} Q_{ji}^{fc}) - 0.5 d_j (Q_{j,s}^{dc} + \sum_{i \in P} Q_{ji}^{fc})^2] \quad (14)$$

$$s.t. \; Q_{j,s}^{dc} = \sum_{i \in P_{j(i)}} Q_{i,s}^{dp} + z_{j,j,s} \quad (15)$$

$$\lambda_{j,s} = c_j - d_j (\sum_{i \in P_{j(i)}} Q_{i,s}^{dp} + \sum_{i \in P} Q_{ji}^{fc} + z_{j,j,s}) \quad \forall s \in S \quad (16)$$

$$\lambda_{j(i),s} = \lambda_{hub,s} + W_{j,s} \quad (\mu_{i,j,s}^{hub}) \quad \forall s \in S \quad (17)$$

$$\sum_{n \in A} z_{j,n,s} = 0 \quad (\mu_{j,s}^{z}) \quad \forall s \in S \quad (18)$$

$$Q_{ji}^{fc} \geq 0 \quad (\mu_{ji}^{fc}) \quad \forall i \in P \quad (19)$$

$z_{j,n,s}$ is the injected power to area $n$ from the hub node from the supplier's viewpoint. The first two terms of the objective function (14) represent the total utility of supplier $j$ for consuming electric energy. The third and fourth terms are the cost of buying energy from the day-ahead market and forward contracts, respectively. Constraint (15) defines the consumed electricity by supplier $j$ in the day-ahead market as the sum of injected power from the grid and produced power by producers in that area. Constraints (16)-(19) are similar to the constraints (9)-(12) but from the viewpoint of the supplier $j$. The suppliers' decision-making variables in forward contracts are $\varepsilon_{ji}^{f} \; \forall i \in P$. Since suppliers are price-takers in the day-ahead market, they do not have decision-making variables in the day-ahead market.

### E. Obtaining the Nash equilibrium of the contracting period

To find the Nash equilibrium of the model, forward contracts equations (1) and (2) and day-ahead market equations (3)-(6) should be solved parallel with optimization problems of all producers (7)-(13) and suppliers (14)-(19). To this end, variables $F_{ij}^{\varepsilon}$ and $Q_{ij}^{\varepsilon}$ are replaced with their equivalents in (1) and (2) at all equations and KKT optimality conditions of all optimizations i.e., (3)-(6) for the TSO, (7)-(13) for all producers, and (14)-(19) for all suppliers are calculated. Moreover, the following market-clearing conditions must be satisfied with market equilibrium:

$$\gamma_{n,s} = x_{i,n,s} = z_{j,n,s} \quad \forall n \in A, j \in C, i \in P, s \in S \quad (20)$$

These equalities ensure that injected power to each node from the viewpoint of TSO, producers, and retailers/big loads is the same. Applying equality (20) causes equations (9) and (16) to get similar for all market players. Hence, these equations can be turned into one equation. A similar process is followed for equations (10) and (17), too. By solving the remained equalities and inequalities of KKT optimal conditions for all market players and the TSO, the Nash equilibrium of the model will be found.

## III. NUMERICAL RESULTS

In this section, the proposed method is applied to the 5-bus PJM test system. The system configuration is presented in Fig. 3. Producers' and suppliers' information is presented in Tables 1 and 2, respectively. The installed capacity of WF1, WF2, and WF3 are 1.5, 2.5, and 2 GW, respectively. The correlation between WF1 and WF2 is 0.7 and the correlation between WF1 and WF3 is 0.85. Sixteen discrete scenarios are generated to model the uncertainty related to the output power of the WFs as shown in Fig. 4. To consider the different concerns of market players in the contracting period about the delivery period, different $\beta_i^p$ and $\beta_j^C$ parameters are assigned to the market players as shown in Table 3 and Table 4. Producer P4 and supplier C3 are assumed to be more concerned about the future than other market players.

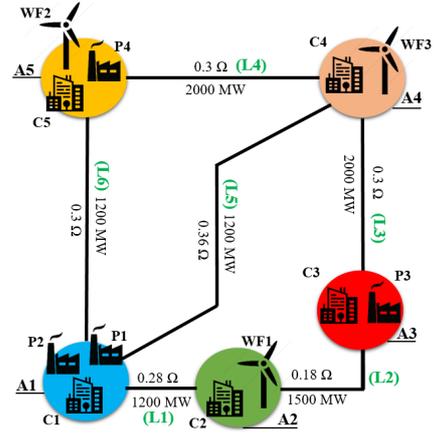

Fig. 3. System configuration

Table 1. Producers cost function parameters

| | Producer number | | | |
|---|---|---|---|---|
| | P1 | P2 | P3 | P4 |
| $a_i$ ($/MWh) | 20 | 16 | 10.8 | 5.6 |
| $b_i$ ($/MW$^2$h) | 0.017 | 0.007 | 0.011 | 0.026 |
| $\overline{Q}_i$ (GW) | 2.5 | 4 | 3.5 | 3 |

Table 2. Suppliers utility function parameters

| | Supplier number | | | | |
|---|---|---|---|---|---|
| | C1 | C2 | C3 | C4 | C5 |
| $c_j$ ($/MWh) | 65 | 61 | 63 | 65 | 66 |
| $d_j$ ($/MW$^2$h) | 0.005 | 0.003 | 0.01 | 0.004 | 0.005 |

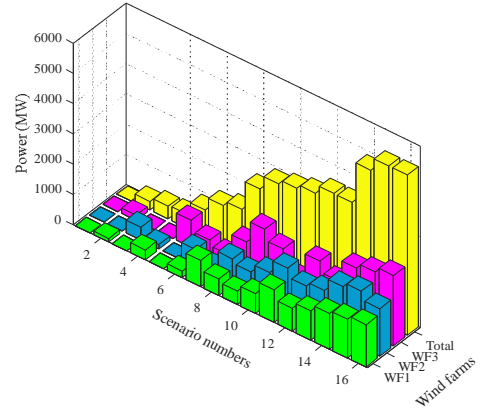

Fig. 4. Uncertainty scenarios for different WFs

Table 3. Coefficients of Exponential PDFs for generating concern scenarios of market players

| | Producer number | | | |
|---|---|---|---|---|
| | P1 | P2 | P3 | P4 |
| $\beta_i^p$ | 0.65 | 0.5 | 0.6 | 0.7 |

Table 4. Coefficients of Exponential PDFs for generating concern scenarios of market players

| | Supplier number | | | | |
|---|---|---|---|---|---|
| | C1 | C2 | C3 | C4 | C5 |
| $\beta_j^C$ | 0.5 | 0.65 | 0.75 | 0.4 | 0.5 |

Table 5. Contract prices and quantities for producers

|  | P1 | P2 | P3 | P4 |
|---|---|---|---|---|
| T. contract quantities (MW) | 1160 | 2091 | 1490 | 820 |
| W.A. of contract prices ($/MWh) | 50.25 | 51.25 | 50.91 | 49.95 |

Table 6. Contract prices and quantities for suppliers

|  | C1 | C2 | C3 | C4 | C5 |
|---|---|---|---|---|---|
| T. contract quantities (MW) | 1047 | 1320 | 739 | 1321 | 1151 |
| W.A. of contract prices ($/MWh) | 50.73 | 50.39 | 51.24 | 50.76 | 50.87 |

### A. Simulation results

To avoid the complexity in the presentation, the sum of the contracts of each producer/supplier with all suppliers/producers and the weighted average (W.A) of the contract prices of each producer/supplier is considered as the total contract quantity and price for each producer/supplier, respectively. Total contract quantities and weighted average contract prices of the market players are presented in Table 5 and Table 6. The weighted average price of all contracts for producers and suppliers is equal to 50.75 $/MWh. The total scheduled power of producers and suppliers in different scenarios are presented in Fig. 5 and Fig. 6. Day-ahead market prices at different areas and transmission system loading are illustrated in Fig. 7. Based on the results, the behavior of market players is affected by both their concerns and congestion in the system. P1 and P2 are in area A1 which due to the congestion in its connected lines, and competition between P1 and P2 has the lowest day-ahead market prices. So, P1 and P2 sell more than 50% of their power in forward contracts with higher prices. P1 is more concerned than P2 about the delivery period and hence, sells a higher percentage of his scheduled power in forward contracts. On the other hand, supplier C1 in area A1 benefits from low day-ahead market prices and buys most of his demand from the day-ahead market. Congestion in the grid also leads to higher average day-ahead market prices in Area A3. This encourages P3 to sell most of his power in the day-ahead market and encourages the C3 to buy most of his required

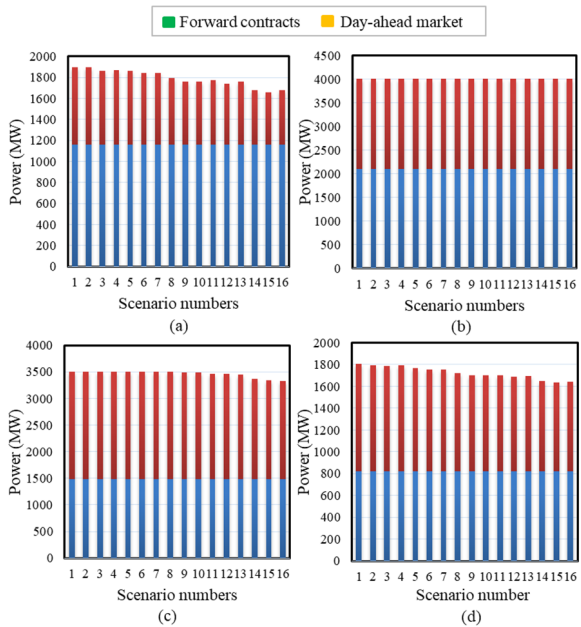

Fig.5. Quantities of forward contracts and scheduled powers in day-ahead market in different scenarios for a) P1, b) P2, c) P3, and D) P4.

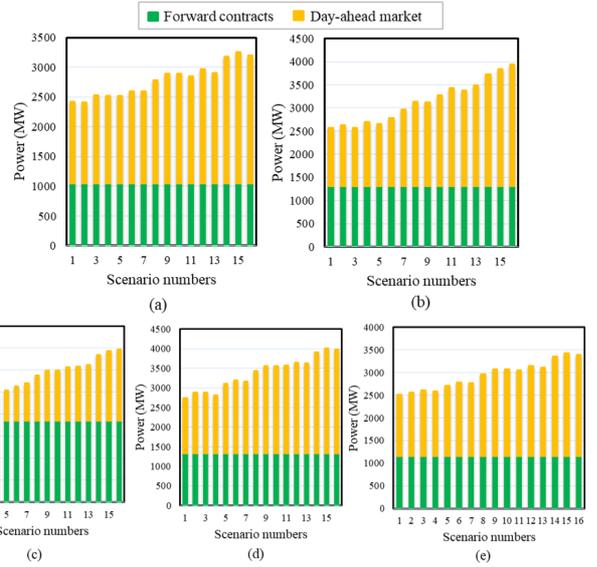

Fig.6. Quantities of forward contracts and scheduled powers in the day-ahead market in different scenarios for a) C1, b) C2, c) C3, D) C4 and e) C5.

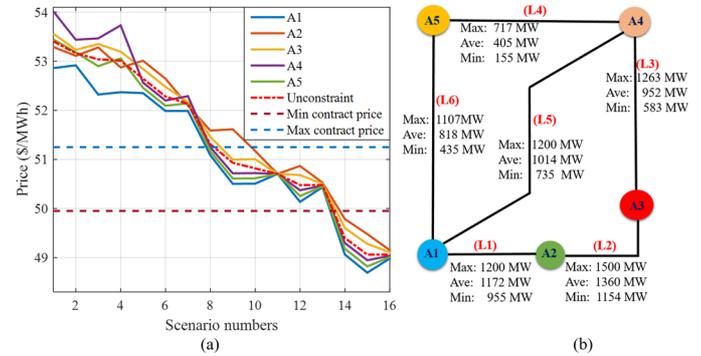

Fig. 7. a) day-ahead market prices, b) transmission system loading

power in forward contracts. As shown in Fig. 7, contract prices are less than the day-ahead market prices in most of the scenarios. In fact, the maximum forward contract price which is 51.25 $/MWh is lower than the average day-ahead market price which is 51.37 $/MWh.

### B. Impacts of grid congestion on forward contracts

To study the impacts of congestion on forward contracts, a modified case is introduced in which the power transmission capacity of the line decreases by 30%. Simulation results are presented in Table 7 and Table 8. LMPs in different areas are also presented in Fig. 8. As shown in Fig. 8, decreasing the line capacities reduces the market prices in areas A1 and A3 considerably. This forces P1, P2, and P3 to sign more forward contracts and encourages A1 and A3 not to involve in forward contracts and benefit from day-ahead market prices as confirmed in Table 7 and Table 8. Due to an increase in the day-ahead market prices in other areas suppliers C2, C4, and C5 are involved in more forward contracts to avoid the risk of high day-ahead market prices.

Table 7. Comparing the ratio of forward contract power quantities to the total scheduled power of producers in two cases

|  | P1 | P2 | P3 | P4 |
|---|---|---|---|---|
| Proposed method | 0.65 | 0.53 | 0.43 | 0.48 |
| Modified case | 0.76 | 0.61 | 0.48 | 0.50 |

Table 8. Comparing the ratio of forward contract power quantities to the total scheduled power of suppliers in two cases

|  | C1 | C2 | C3 | C4 | C5 |
|---|---|---|---|---|---|
| Proposed method | 0.37 | 0.42 | 0.64 | 0.39 | 0.39 |
| Modified case | 0.18 | 0.75 | 0.26 | 0.68 | 0.46 |

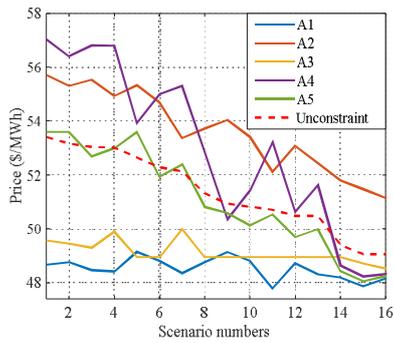

Fig. 8. The day-ahead market price in different areas at different scenarios in the modified case

IV. CONCLUSIONS

In this paper, the impacts of transmission system constraints and large-scale integration of WPPs on the behavior of the market players in the forward contract and day-ahead electricity market are studied. The day-ahead electricity market and forward contracts have mutual impacts on each other. So, to study the behavior of the market players in the system, both contracts and the day-ahead market and their interactions should be considered. To this end, a Cournot Nash equilibrium model was presented for the forward contract negotiation period considering different possible outcomes for the day-ahead electricity market due to the wind power uncertainty and transmission system constraints and mutual impacts of forward contracts and the day-ahead market.

Simulation results show that the behavior of the market players in the system differs depending on their concerns about the delivery period and their location in the grid. In similar conditions, as the concern of market players about the future increases they contract more power in the system. In the areas where there is a greater probability of reducing the market price the quantities of contract powers for producers are increased and the quantities of contract powers for consumers in those areas are reduced. Simulation results also indicate that increasing the congestion in the grid affects the market players differently based on the impacts of congestion on the estimated day-ahead market prices in the delivery period.

ACKNOWLEDGMENT

This work emanated from research conducted with the financial support of the Flexible Energy Denmark (FED) project.